\documentclass[12pt,twoside]{article}

\evensidemargin=\oddsidemargin
\def\Title#1#2#3{%
    \baselineskip=18pt
    \begin{center}
          {\large\bf\uppercase{#1} \\ }
          \bigskip\bigskip
          {#2} \\
          {#3} \\
    \end{center}}
\newcommand{\Abstract}[1]{\begin{abstract}#1\end{abstract}}

\begin{document}

\vspace*{1cm}

\Title{QUANTUM GEOMETRODYNAMICS\\
CREATES NEW PROBLEMS}%
{T. P. Shestakova}%
{Department of Theoretical and Computational Physics,
Southern Federal University\footnote{former Rostov State University},\\
Sorge St. 5, Rostov-on-Don 344090, Russia \\
E-mail: {\tt shestakova@phys.rsu.ru}}

\vspace{1cm}

\Abstract{The development of last years in quantum geometrodynamics
highlights new problems which were not obvious in its first
formulation proposed by Wheeler and DeWitt. At the first stage the
main task was to apply known quantization schemes to gravitational
field or a certain cosmological model. This way has led to the
realization of the fact that a quantum description of the Universe
is impossible without implicit or explicit indication to a
reference frame presented by some medium, filling the whole
Universe, with its own equation of state and thermodynamical
properties.

Thus the questions arise, should one seek for a ``privileged''
reference frame or consider all the variety of gauge conditions and
appropriate solutions for the wave function? It is worth noting
that thermodynamical properties of such a quantum Universe would
also depend on a chosen reference frame to some extent. So, we need
a self-consistent quantum theoretical and thermodynamical
description of the Universe.}

\vspace{1cm}

In this talk I would like to present my point of view on the
problems of quantum geometrodynamics, as they can be seen now,
forty years after the first significant attempt to construct full
quantum theory of gravity made by DeWitt in 1967 \cite{DeWitt}.
As all of us well know, first attempts to apply quantum theory
to gravitational field immediately faced enormous obstacles, and
the aim of this my talk is to demonstrate the interrelation
between those obstacles and the problems quantum geometrodynamics
meets now.

At the first stage the main task was to apply known quantization
schemes to gravitational field or a certain cosmological model. It
was realized that the main difficulty consisted in the nature of
general relativity as a completely covariant theory that ran
counter to efforts to build a Hamiltonian formulation of it as the
first step on the way of its quantization. The difficulty was
referred to as ``the problem of constraints''. Meanwhile, in 1950s
Dirac published his outline of a general Hamiltonian theory
\cite{Dirac1,Dirac2} which was in principle applicable to any
system with constraints, in particular, to gravitational field. The
next important step was done by Arnowitt, Deser and Misner
\cite{ADM} who proposed a special parametrization of gravitational
variables that made the construction of Hamiltonian formalism
easier and admitted a clear interpretation. The third source of
DeWitt theory was the ideas of Wheeler concerning a wave functional
describing a state of gravitational field \cite{Wheeler1,Wheeler2}.
Dirac approach to quantization of systems with constraints,
Arnowitt -- Deser -- Misner (ADM) parametrization and Wheeler ideas
are the three cornerstones on which the Wheeler -- DeWitt quantum
geometrodynamics is based.

It seemed that the initial obstacles had been overcome. However,
the Wheeler -- DeWitt quantum geometrodynamics encountered a number
of fundamental problems which cannot be resolved in its own limits
and which have made a way for its strong criticism. So, Isham
\cite{Isham} wrote: ``...although it may be heretical to suggest
it, the Wheeler -­ DeWitt equation -- elegant though it be
-- may be completely the wrong way of formulating a quantum theory
of gravity''.

What did make Isham to claim it? There exist serious doubts that
the Dirac approach can be applied to gravitational field. The
central part in the Dirac approach is given to a postulate,
according to which each constraint $\varphi_m(q,p)=0$ after
quantization becomes a condition on a state vector, or wave
functional, $\Psi$: $\varphi_m\Psi=0$.
Let us emphasize that it is indeed a postulate, since it cannot be
justified by the reference to the correspondence principle. The
role prescribed to the constraints could be explained by the fact
that at the classical level, the constraints express gauge
invariance of the theory. It was initially believed that imposing
constraints at the quantum level would also ensure gauge invariance
of wave functional. But what grounds do we have to expect it?
Strictly speaking, the founders of quantum geometrodynamics have
not investigated this issue and gauge invariance of the theory has
not been proved. It leads us to the next fundamental problem: Could
we consider quantum geometrodynamics as a gauge-invariant theory?

An important role was played by the ADM parametrization: it is the
ADM parametrization that enables one to write gravitational
constraints in the form independent of gauge variables -- the lapse
and shift functions $N$, $N_i$. It gave rise to an illusion that
the theory in which the main equations are those of constraints
must not depend on a choice of gauge conditions. At the same time
the ADM parametrization introduces in 4-dimensional spacetime a set
of 3-dimensional hypersurfaces (the so-called (3+1)-splitting). But
fixing (3+1)-splitting prescribes particular values for the lapse
and shift functions \cite{MM1,MM2} that is equivalent to fix a
reference frame, and gauge invariance breaks down. Thus, the
Hamiltonian constraint loses its sense and, with the latter, so
does the whole procedure of quantization.

The third point was the idea by Wheeler that the wave functional
must be determined on the superspace of all possible 3-geometries
However, the statement that the wave function must depend only on
3-geometry is just a declaration without any mathematical
realization. As we know, the state vector always depends on a
concrete form of the metric.

Gauge invariance of the Wheeler -- DeWitt theory can hardly be
proved or refute within canonical quantization approach. Path
integration approach is more powerful, since in this case gauge
invariance of the path integral, and the theory as a whole, is
ensured by asymptotic boundary conditions. In first works devoted
to derivation of the Wheeler -- DeWitt equation from the path
integral \cite{BP,Hall}, asymptotic boundary conditions were
tacitly adopted without careful consideration if they are
justified. In the works of our group in the end of 1990s attention
was focused on the circumstance that the Universe is topologically
non-trivial system without asymptotic states \cite{SSV1,SSV2}. It
led to the conclusion that the picture of quantum evolution of the
Universe cannot be independent of a reference frame in which this
evolution is studied.

Our work were not the only works in this trend. I should mention
the pioneer paper by Brown and Kucha\v r \cite{BK}, the works by
the group of Montani and collaborators \cite{MM1,MM2,BM} and others
that opened the way to ``Evolutionary Quantum Gravity''. In all the
approaches a reference frame is presented by some medium, filling
the whole Universe, with its own equation of state and
thermodynamical properties. The reference frame is introduced in
these approaches by different ways. In the approach by Brown and
Kucha\v r the reference frame is related with incoherent dust, In
the work by Montani and collaborators it is done by means the
so-called kinematical action. In the extended phase space approach
it is argued that any gauge-fixing term in Batalin -- Vilkovisky
(Faddeev -- Popov) effective action describes a medium with
mentioned above properties.

Returning to the problem of constraints, one should confess that it
has not been solved in the sense that we failed to construct a
gauge-invariant quantum theory of gravity for the whole Universe.
Instead of a gauge-invariant theory, in a mathematically consistent
approach one should reject the Wheeler -- DeWitt equation as a
quantum version of the Hamiltonian constraint, and reestablish the
role which Schr\"odinger equation plays in any quantum theory and
which it lost in quantum geometrodynamics. A quantum
geometrodynamical Schr\"odinger equation appears to be
gauge-dependent. Any changing a reference frame results in varying
gauge conditions and, in its turn, in modifying of the
Schr\"odinger equation for a wave function of the
Universe, as well as its solution. It can be expressed by the
following scheme:

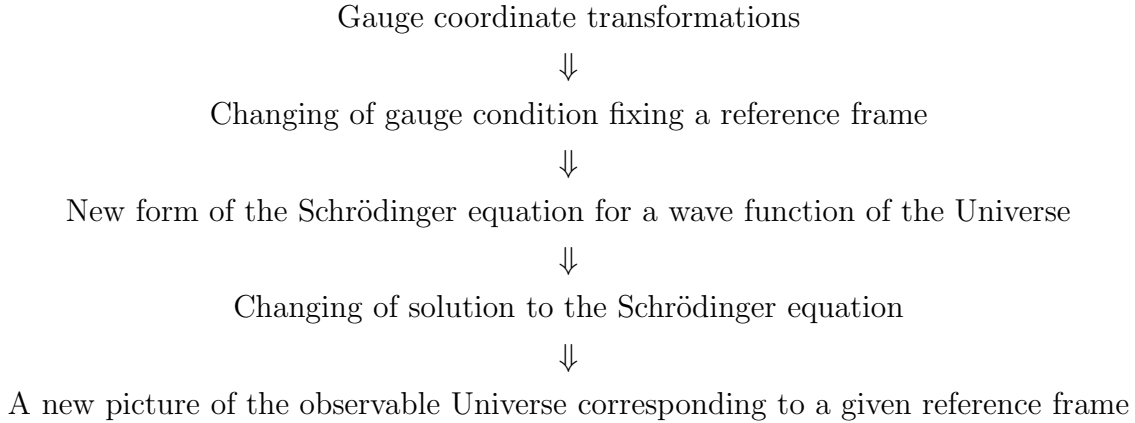
\begin{center}
Gauge coordinate transformations
\par $\Downarrow$
\par Changing of gauge condition fixing a reference frame
\par $\Downarrow$
\par New form of the Schr\"odinger equation
for a wave function of the Universe
\par $\Downarrow$
\par Changing of solution to the Schr\"odinger equation
\par $\Downarrow$
\par A new picture of the observable Universe
corresponding to a given reference frame
\end{center}

Let us try to give an assessment of the modern state of quantum
geometrodynamics. The first possibility is to search for arguments
in favor of some privileged reference frame in which the picture of
the Universe evolution would better correspond with observational
data. For example, a significant physical argument could be if
quantum evolution of the Early Universe would ensure for inflation
stage and further classical evolution of the observable Universe.
However, today we do not have available such significant arguments
and do not have any grounds to postulate a privileged reference
frame. In my opinion, we should consider all possible situations
and feel about for relations between classes of gauge
transformation of diffeomorphism group and classes of solutions to
the Schr\"odinger equation. This task is very laborious since the
structure of diffeomorphism group is known to be very complicated.
Nevertheless, one can start, as usual, from well-studied subgroups
and try to find the way.

In a full description one should also take into account
thermodynamical properties of a quantum Universe filled with a
medium playing the role of a reference frame. Indeed, one of
possible methods to build thermodynamics of the system under
consideration is to write a density matrix through a path integral
with Euclidean version of an action (in our case it is a gauged
gravitational action), so that thermodynamical properties of the
system would depend on a chosen reference frame as well. It must
not be surprising for us, since the example of Rindler space
teaches us that thermodynamical properties could actually change
after going over to another frame. But we yet need a clear
interpretation of quantum gravitational phenomena taking place
under this transition.

And the question remains: Do we have to appeal to an underlying
theory to solve these problems or can we hope to resolve them
within the framework of quantum geometrodynamics itself? We are
still far from the so-called final theory of Everything. In our
attempts to approach the final theory we rely on our experience in
ordinary quantum field theory and quantum gravity, and often face
the same problems. So, I do not believe that we should give up the
search for better understanding the principles on which quantum
geometrodynamics ought to be grounded.

\end{document}